# An improved spectrophotometry tests the Einstein-Smoluchowski equation: a revisit and update

*Jiangbo (Tim) Zhao[†]\*, Cong Qi[†], Guangrui Li[†], and Markus A. Schmidt[†,‡,⊥]*

[†] Leibniz Institute of Photonic Technology, Albert-Einstein-Straße 9, 07745 Jena, Germany

[‡] Abbe Center of Photonic and Faculty of Physics, Friedrich-Schiller-University Jena, Max-Wien-Platz 1, Jena 07743, Germany

[⊥] Otto Schott Institute of Material Research, Fraunhoferstr. 6, 07743 Jena, Germany

E-mail: jiangbo.zhao@leibniz-ipht.de

**ABSTRACT:** Theoretical prediction and experimental measure of light attenuation in chemically pure and optically transparent solvents have attracted continuous attention, due in part to the curiosity to nature, and in part to increasing calls from solvent-related applications. Yet hitherto, a majority of accurate spectrophotometric measurements of transparent solvents upon visible light radiation often end up using long-path-length cells, usually over dozens of cm, rendering the measure costly and complex; meanwhile, the guidance on choosing the Einstein-Smoluchowski equation or its variants as the best formula to predict the light scattering in solvent has remained elusive. Here we demonstrate a simple, versatile and cost-effective spectrophotometric method, enabling sensitivity $10^{-4}$ dB/cm with over 0.5 cm differential path length based on using standard double-beam spectrophotometre. We prove the method reduces the path length by a factor of 100 while still making its closest to the record-low measurement of solvent extinctions. We also validate all the present equations used for predicting the light scattering in solvent possess the similar capacities, suggesting that the criterion for the choice of the appropriate formula simply depends on the equation's practicability. Following the elucidation of the wavelengths range where the light scattering dominates the extinction, we further identify differences between scattering coefficients via the theoretical predictions and experimental measures, exposing the need for an improved theory to account for the solvent scattering phenomenon.

**TOC GRAPHICS**

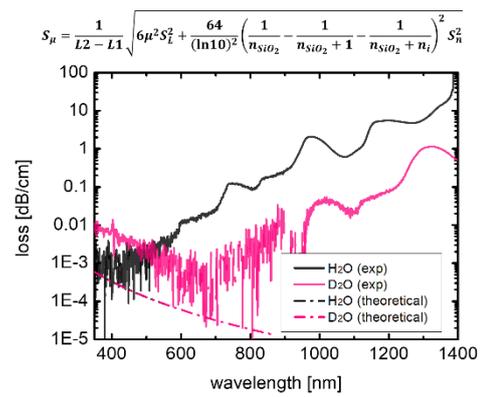

$$S_H = \frac{1}{L2 - L1}\sqrt{6\mu^2 S_L^2 + \frac{64}{(\ln 10)^2}\left(\frac{1}{n_{SiO_2}} - \frac{1}{n_{SiO_2} + 1} - \frac{1}{n_{SiO_2} + n_t}\right)^2 S_n^2}$$

We develop a highly accessible and accurate spectrophotometry, and identify shortcomings of the Einstein-Smoluchowski equations in solvent scattering prediction.

## 1. Introduction

Attenuation of light over the ultraviolet (UV) to mid-infrared (MIR) wavelengths range in chemically pure and optically transparent solvents is caused primarily by absorption and scattering. The absorptions refer to the electronic transitions and vibrational resonances, related with the electronic and atomic polarizations in molecules. [1-4] The scattering stems from the density fluctuation of solvent (i.e., mean square fluctuation of the dielectric constant). Its theoretical prediction, which has attracted extensive interests, including many eminent scientists such as Rayleigh, [5] Smoluchowski, [6] Einstein, [7] Raman, [8] Debye, [9] Oppenheimer, [10] etc, is initially undertaken by executing the Einstein-Smoluchowski equation, and since 1960s is conducted mainly through the equation's variant forms. However, up until now a consensus on the best formula for predicting the magnitude of solvent scattering has not been reached. [11, 12]

Besides theoretical understanding of how light interacts with solvent, experimental measure of the corresponding coefficients, e.g., extinction coefficient ($\mu$), absorption coefficient ($a$) and/or scattering coefficient ($b$), is of equal significance. Firstly, high-accuracy empirical coefficients can be used to examine the validity of the theorem in predicting the light-solvent interaction. Secondly, the pursuit of accurate coefficients mutually couples with the advancement of a series of technologies, including the production of high-quality glass and high-purity solvent, and deployment of advanced optical systems and techniques, such as spectrophotometre adiabatic laser calorimetre, optoacoustic spectroscope and integrating cavity absorption metre. [13, 14] Thirdly, the coefficients, as a kind of physical constant, are essential to fields as diverse as lidar bathymetry, neutrino observation detector construction, photosynthetic process, optofluidic device development and microscopy imaging medium. [13, 15-21] Despite enormous success in measuring the solvent coefficients, especially over the visible spectral region such as for $H_2O$, their conduct has had to use long-path-length cells, at least dozens of cm, [4, 10-14, 22-25] otherwise, large errors appear around the extinction minima (Table S1). As seen in industry for spectrophotometric evaluation of the materiality of ultrapure $H_2O$, a modified instrument accommodating a 30 cm-length cell and auxiliary integrating sphere is necessarily deployed for the measurement. [26]

In this work, we develop a convenient, versatile and cost-effective spectrophotometry allowing for accurate measurements of regular and deuterated solvents over wavelengths 350-1400 nm, based on using standard double-beam spectrophotometre with 0.5 cm differential path length, not bothering to modify the instrument (Section 2). Since the absorption and scattering contribute to the solvent extinctions in varying proportions over the wavelengths concerned, we follow determining the spectral division where the extinction is primarily contributed by light scattering (Section 3.1). We then confirm that the Einstein-Smoluchowski equation and its variants are almost the same in theoretical predictions of the magnitude of solvent scattering, and further demonstrate the shortcomings of the present equations (and theory) in comparison with the accurate measured results, suggesting the need for better comprehension of the science underlying light scattering in solvent (Section 3.2). We summarise the outcomes in Section 4.

## 2. An improved spectrophotometry

### 2.1 Optical paths of liquid-filled and empty cuvettes

In general, for either liquid-filled or empty cuvettes, the total transmitted light $I_t$ contains the flux traversing the quartz glass and medium for once, thrice, and so on, during which the reflections and refractions take place at the air-glass and/or liquid-glass interfaces (Figure S3). The reflections and transmissions, for an interface, are described by Fresnel equations $R_{i,\mathrm{SiO_2}} = (\frac{n_{\mathrm{SiO_2}} - n_i}{n_{\mathrm{SiO_2}} + n_i})^2$ and $T_{i,\mathrm{SiO_2}} = 1 - R_{i,\mathrm{SiO_2}}$, where $n_{\mathrm{SiO_2}}$ and $n_i$ are the refractive indices of the cuvette and medium, respectively. In accordance with the Beer-Lambert law (Section S2.1), the transmissions (i.e., extinctions) of the quartz glass and medium are expressed as $T_{\text{SiO}_2} = 10^{-\mu_{\text{SiO}_2} L_{\text{SiO}_2}}$ and $T_i = 10^{-\mu_i L_i}$, where $\mu_{\text{SiO}_2}$ (and $\mu_i$) and $L_{\text{SiO}_2}$ (and $L_i$) are the extinction coefficients and path lengths.

For liquid-filled cuvette, the refractive indices of various solvents over different wavelengths are generally within 1.3-1.5, the magnitude of the reflection $R_{i,\text{SiO}_2}$ at the liquid-glass interface thus varies in between $10^{-3}$ and $10^{-5}$. Given inconsiderable $R_{i,\text{SiO}_2}$, the incident light $I_0$, after a reflection at the first air-glass interface with a fraction of $R_{air,\text{SiO}_2}$, undergoes an attenuation by a factor $10^{-(2\mu_{\text{SiO}_2} L_{\text{SiO}_2} + \mu_i L_i)}$ till the second air-glass interface, from which an amount of $I_0 (1 - R_{air,\text{SiO}_2})^2 10^{-(2\mu_{\text{SiO}_2} L_{\text{SiO}_2} + \mu_i L_i)}$ exits as the first-order transmitted light, and the rest is reflected and subjected to the iterative reflections and transmissions as described above (see details in Figure S3a). The total of the different-order transmitted light is summed as (the deduction process in Section S2.2):

$$I_{t,f} = I_0 \big(T_{air,\text{SiO}_2} T_{\text{SiO}_2}\big)^2 T_i (1 + \delta_f), \tag{1}$$

where $\delta_f$, defined as the ratio of a total amount of the higher-order transmitted light to the first-order counterpart, is approximately equal to $0.001 T_i^2$. Since $T_i$ is a function of the wavelength, $\delta_f$ exhibits the spectral distribution characteristic.

The Eq. (1) suggests that, with the measured ratio $I_{t,f}/I_0$, solvent extinction $T_i$ (or $\mu_i$) can be extracted basically via three different strategies: (1) substituting the exact numeric values of $T_{air,\text{SiO}_2} T_{\text{SiO}_2}$ into the equation; (2) increasing the path length of a cuvette to the extent that the extinction of the filled liquid is far more than that by product of $T_{air,\text{SiO}_2} T_{\text{SiO}_2}$, hence making the magnitude of $T_{air,\text{SiO}_2} T_{\text{SiO}_2}$ be reduced to approximately one, and then $T_i$ equivalent to $I_{t,f}/I_0$; (3) or performing multiple spectrophotometric scans for different-length cuvettes in the empty and filled states, respectively, thus being able to cancel the cuvette-related terms and to derive $T_i$. One of the sufficient conditions for executing the strategy three, obviously, is that the glass walls between the cuvettes are as much identical as possible.

With regard to the spectrophotometric measurement of transparent solvents, even if the strategy one is possible, the determination of high-accuracy $T_{SiO_2}$ and $T_{air,SiO_2}$ for each glass walls and faces (including interior and exterior glass surfaces) can be highly laborious. The other two strategies are therefore often implemented, represented by the so-called One- and Two-step methods (bottom and middle panels in Figure 1) accordingly. Though the One-step method, based on the strategy two, has recorded the lowest extinction measurement, [13] this methodology demands use of dozen-of-cm-length glass cell, configuration of customised spetrophotometre, installation of auxiliary apparatus and fill of large-volume solvent, [26] all of which inherently contradict with the goal of developing a convenient, versatile and cost-effective spectrophotometry.

Under this circumstance the strategy three seems left as the only option to be exploited. The optical paths of the empty cuvette used for the measurement are thus appreciated. As shown in Figure S3b, iterations of light attenuation over the glass walls and reflections at each of the four air-glass interfaces are involved, leading to first-order and a series of higher-order transmitted light. The summation of the total transmitted light is given as (the deduction process in Section S2.2):

$$I_{t,e} = I_0 \big(T_{air,SiO_2}^2 T_{SiO_2}\big)^2 T_{air}(1 + \delta_e), \qquad (2)$$

where $\delta_e$, varying between 0.0063 and 0.0078 over the wavelengths concerned, is viewed as a wavelength-insensitive item, approximately 0.007.

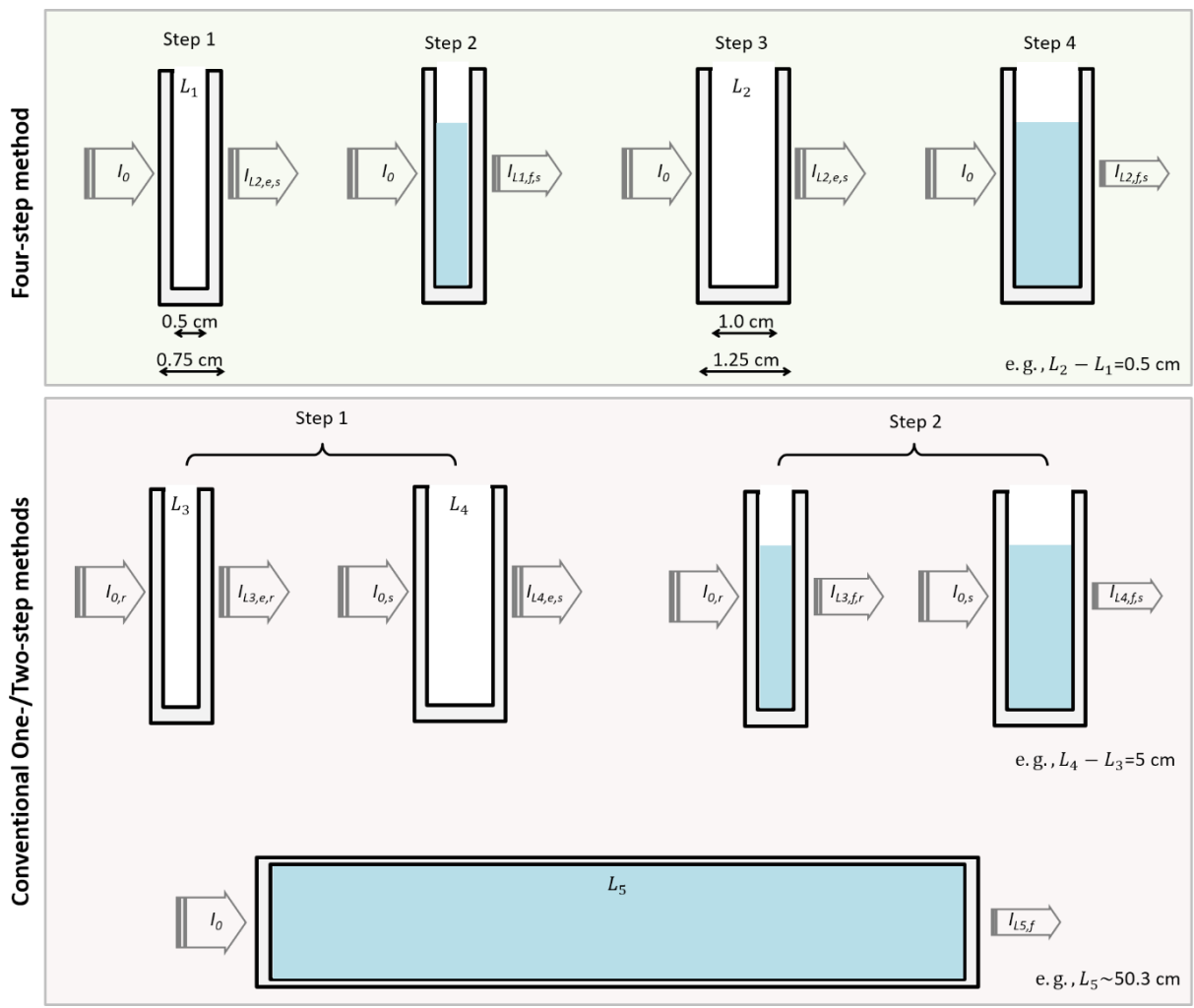

**Figure. 1** The flowchart of the Four-step method (top panel) and conventional One- and Two-step approaches (bottom and middle panels). The subscript annotations denote the cuvettes ($e/f$ – under empty and filled states), in reference and sample compartments ($r/s$), with different path lengths ($L_{1/2}$). For instance, incident and transmitted light in the Four-step method are described by $I_0$, $I_{L1,e,s}$, $I_{L1,f,s}$, $I_{L2,e,s}$ and $I_{L2,f,s}$, respectively, and varied-width arrows in the scheme are used to illustrate the attenuation of radiant fluxes at a different extent after a medium.

## 2.2 Proposal of the Four-step method

The strategy three is, thus far, executed merely by implementing the Two-step method (Figure 1, middle panel), which however has suffered from large errors as opposed to the benchmark extinction coefficients measured by the One-step method, as manifested in Table S1. The crux of the Two-step method lies in the simultaneous use of a pair of cuvettes in the reference and sample compartments. As elaborated in Sections 2.3, S1.5 and S3.3, for the Two-step method, the dual light beams in a spectrophotometre differ from each other, and complete cancellation of the cuvette-related terms is not possible, even for cuvettes made up of the exactly same glass walls. To better execute the strategy three, we propose the "Four-step method" (top panel in Figure 1, see details in Section S3.1). In short, rather than using a pair of cuvettes simultaneously, the method only conscripts the sample compartment into the sequential measure of two-different-path-length cuvettes in the empty and filled states, obtaining absorbance $A_{L1,e,s}$, $A_{L1,f,s}$, $A_{L2,e,s}$ and $A_{L2,f,s}$, respectively. The solvent extinction coefficient, based on the Four-step method, is written as (the deduction process in Section S3.2):

$$\mu = \frac{\Delta A_{L2,s} - \Delta A_{L1,s}}{L_2 - L_1} + \frac{Y}{L_2 - L_1}, \qquad (3)$$

where $\Delta A_{L1,s} = A_{L1,f,s} - A_{L1,e,s}$, $\Delta A_{L2,s} = A_{L2,f,s} - A_{L2,e,s}$, and $Y$ is described as:

$$Y = Log_{10}\left(\frac{1 + \delta_{L2,f,s}}{1 + \delta_{L1,f,s}}\frac{1 + \delta_{L1,e,s}}{1 + \delta_{L2,e,s}}\right). \qquad (4)$$

As discussed in Section S3.2, as long as the pair of the cuvettes would appear nearly identical, $Y$ is trivial and negligible.

*2.3 Theoretical validation of the Four-step method*

The measuring capacity of the Four-step method is theoretically assessed by compiling its systemic and random (measurement) errors. For the sake for convenience, only the transverse electric (TE) mode is analysed, and the transverse magnetic (TM) mode is not separately considered due to its high similarity with TE mode.

Light beam and glass cuvette both are never perfect. The so-called collimated beam passing through the instrument compartment is in actuality a paraxial ray, which can be subjected to further lateral displacement and diameter change with inserting even ideal cuvette. The faulty characteristics of the cuvette, such as unevenness, nonparallelism, etc, would additionally increase non-orthogonality angle between the incident light and cuvette, distorting the beam further. Hence the insert of cuvette in the compartment perturbs the light beam in a way that differs from those used for the baseline correction scan, causing the systemic errors that cannot be avoided.

Physical reasons responsible for the systemic errors suggest the superiority of the Four-step method over the Two-step method. In operation of standard double-beam spectrophotometre, light beams traversing the reference and sample compartments are highly similar but not exactly the same. Any pair of cuvettes are not identical and never will be. The concurrent use of such pair, as the Two-step method does, will multiply the disparity between the beams, dilating the systemic errors, as discussed in Section S3.3 and evident in Table S1. Rather than perturbs two light beams as the Two-step method did, the Four-step method only interferes with the beam in the sample compartment where the cuvette is placed, and the beam in the reference channel is not bothered from the baseline correction throughout the sample measurement since only the air is there all the time. The seemingly small operation change conducted in the Four-step method is able to effectively diminish the disparity between the beams, then the bias upon the detector, allowing to suppress the systemic errors at the most extent.

The generation of random (measurement) errors in the Four-step method is due in part to the non-identical cuvettes and in part to the non-orthogonality angle between the cuvette and light beam. Such that the random errors are a function of multiple variables, including the (real) refractive index of cuvette ($n_{SiO_2}$), the angle of incidence at an interface ($\theta$) and the length of cuvette ($L_i$). A propagation of error of these variables formulates the magnitude of the random errors for the extinction coefficient, which is expressed as (the deduction process in Section S4.1):

$$S_\mu = \frac{1}{L2-L1} \sqrt{6\mu^2 S_L^2 + \frac{64}{(\ln 10)^2} \left( \frac{1}{n_{SiO_2}} - \frac{1}{n_{SiO_2}+1} - \frac{1}{n_{SiO_2}+n_i} \right)^2 S_n^2}. \qquad (5)$$

Eq. (5) reveals that $S_\mu$ is determined by the solvent- and cuvette-related variables ($n_i$ and $\mu$, and $n_{SiO_2}$ and $L_i$), as well by their uncertainties ($S_L$ and $S_n$), but little responsive to the change of angle $\theta$. As shown in Figure 2a, the angle-induced relative errors are capped at 0.185% and 0.055% for solvent with refractive index 1.3, and detuned to 0.172% and 0.050% with increasing the refractive index to 1.5. Since the wavelength-dependent refractive index for any pure solvents in the wavelengths range concerned generally varies between 1.3 and 1.5, the errors associated with the angle are negligible.

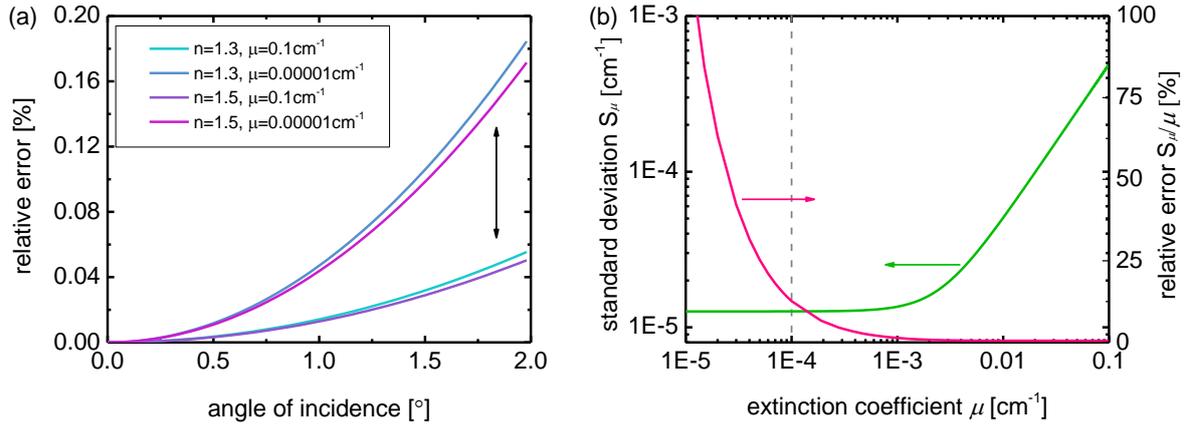

**Figure. 2** (a) The relative errors as a function of the angle of incidence (0-2°) for a pseudo solvent with extinction coefficients in the range of 0.00001- 0.1 cm⁻¹ (refractive indices in between 1.3 and 1.5). Note that in practice the upper bound of the angle between the cuvette and incident light is about 2.0°. (b) The standard derivation $S_\mu$ (SD, green curve for left axis) and relative errors $S_\mu/\mu$ (pink curve for right axis) for a pseudo solvent as a function of the extinction coefficients from 0.00001 to 0.1 cm⁻¹ (differential length 0.5 cm), obtained in accordance with Eq. (6). The vertical dash line at 0.0001 cm⁻¹ defines the lower bound extinction coefficient that a majority of solvents could approach (except $H_2O$ and $D_2O$).

As specified by the manufacturer, the thickness deviation of the glass wall is ± 0.001 cm, so does the cuvette $L_i$; the $n_{SiO_2}$ deviation of Suprasil® quartz glass is ± 3×10⁻⁵;[27] and the nominal differential path length between cuvettes is 0.5 cm. For the current experimental system, $S_\mu$ is rewritten as:

$$S_\mu < 2\sqrt{0.00004 + 6\mu^2} \cdot 10^{-3}. \tag{6}$$

As shown in Figure 2b (green colour curve), $S_\mu$ staggers around $10^{-5}$ cm$^{-1}$ for $\mu$ below 0.001 cm$^{-1}$; and as indicated by the pink colour curve, $S_\mu/\mu$ resides at about 12.6%, 1.3%, and 0.1% for the extinction coefficients at 0.0001, 0.001 and 0.1 cm$^{-1}$, respectively. These analysis suggests that, by only using 0.5 cm differential length, the Four-step method allows for high-accuracy measurements, and if the length were extended to 5 cm, the measurement errors over a wide range of wavelengths would fall to within 3% for any solvents.

### 2.4 Experimental validation of the Four-step method

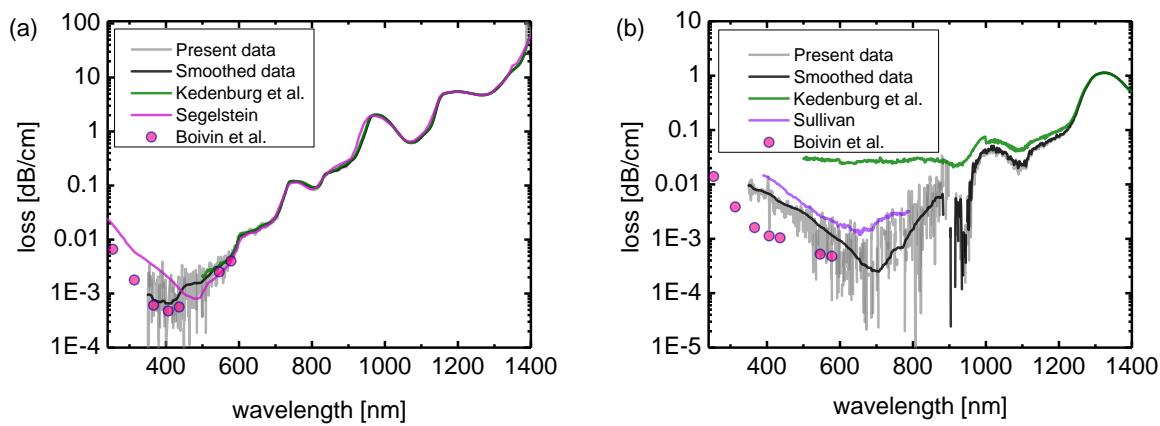

**Figure. 3** The extinction coefficients of two notable weakly absorbing solvents $H_2O$ (a) and $D_2O$ (b) over the wavelengths 350-1400 nm. The grey and black lines stand for the measured (at room temperature) and smoothed data (via 50-60th percentile) in this work. The other curves refer to the data extracted from the publications as noted. On a logarithmic scale the trivial minus values from 850 to 950 nm in extinction of $D_2O$ result in the disruption of the curve, ascribed to that the instrument noise makes the extinction volatile in that section.

The viability of the Four-step method is also assessed by comparing the experimentally measured extinction coefficients of $H_2O$ and $D_2O$, between this method and other approaches.

As shown in Figure 3, compared with the respective benchmark results by Segelstein [28] and by Sullivan [29], the extinction minima measured in this work reduce about -22% for $H_2O$ and -83% for $D_2O$ (Table S1); and as opposed to widely referred publications, [13, 24, 28-30] the position and magnitude of the extinction minima in our research are closest to the record-low extinction coefficients by Boivin *et al*, [13] who, however, accomplished those data through using a 50.3 cm-path-length glass cell in the customised spectrophotometre (Boivin *et al*.: 0.000476 dB/cm at 405 nm for $H_2O$ and 0.000478 dB/cm at 578 nm for $D_2O$, this work: 0.000656 dB/cm at 405 nm for $H_2O$ and 0.0002495 dB/cm at 706 nm for $D_2O$). Compared to the Two-step method measurement for $H_2O$ by using around 1 cm path length, [17] our method gives rise to an improved accuracy by a factor of about 450 with only using 0.5 cm differential path length. Relative to the recent work by Kedenburg *et al*. who used a 4.0 cm cell to measure $D_2O$, [30] our measurement demonstrates an improvement about two orders of magnitude, by using a fairly short differential path length – 0.5 cm.

The theoretical and experimental investigations above have validated the advantage of the Four-step method over the conventional spectrophotometric approaches. As suggested by Figure S4 and by Eq. (6), in addition to in association with the properties of cuvettes and solvents, the detection limit of the Four-step method is affected by the instrument characteristics. To access the best sensitivity for the measurement, two spectrophotometres are employed in this work in a complementary wavelengths range (see Sections S1.4 and S1.5).

## 3. Light scattering in solvent: a revisit and update

### 3.1 Molecular vibrations in solvent

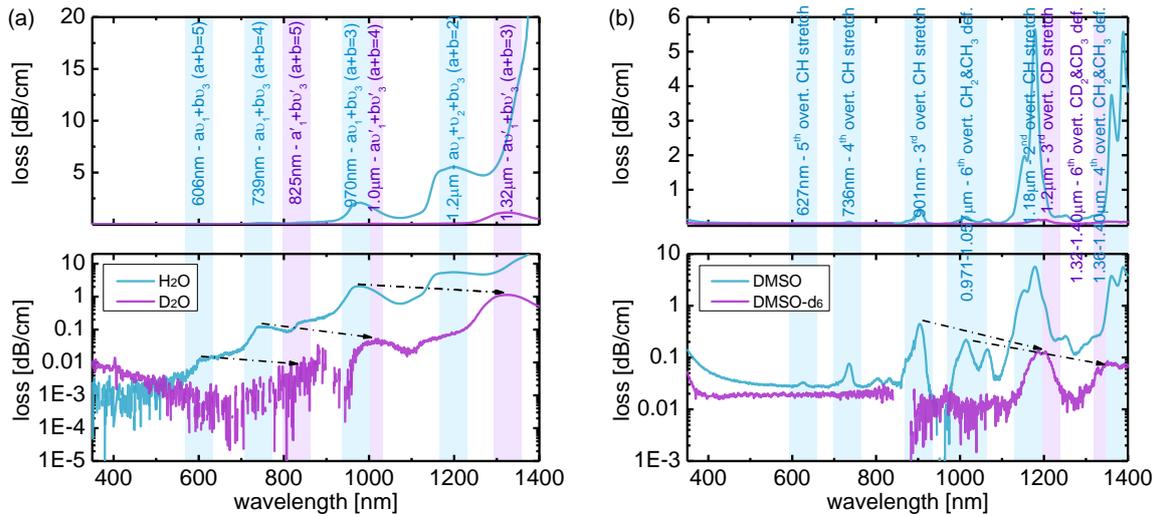

**Figure. 4** Extinction coefficients of (a) $H_2O$ and $D_2O$, and (b) DMSO and DMSO-$d_6$ on the linear (top panels) and logarithmic (bottom panels) scales, measured by the Four-step method. Their overtones and combination bands are denoted by pale blue colour columns for regular solvents, and by light magenta colour columns for deuterated counterparts, with their fundamental vibrations summarised in Section S5. As indicated by the right- and downward-pointing arrows, the elementary substitution of hydrogen by deuterium leads to the reductions in both fundamental resonance frequency, [31] and anharmonicity constant and fundamental band strength, [1] accounting for the red shifts of the resonance wavelengths and the lowering of the magnitude of the same quanta-governed vibrations in deuterated solvents, compared to their protonated counterparts. The fundamental resonance frequency $\upsilon_0 \sim \sqrt{k/m^*}$, where $k$ is the force constant, and $m^*$ is the reduced mass for diatomic bonds $m^* = m_1 m_2/(m_1 + m_2)$. On the logarithmic scales, the broken extinction curves for solvents $D_2O$, DMSO and DMSO-$d_6$ over the wavelengths 850-950 nm again are caused by the instrument noise-perturbed trivial minus values.

The intermixing of scattering and absorptions constitutes the resulting extinction of solvent. Separation of the scattering-originated extinction from that by absorptions is a prerequisite for studying and understanding of light scattering in solvent. That is to determine the wavelengths range where the solvent scattering dictates the extinction rather than the absorptions, including both electronic transitions-induced absorptions and molecular vibrations-related absorptions. Given that $H_2O$ and DMSO exhibit UV cut-off wavelengths at 190 nm and 268 nm, [32] and their deuterated counterparts possess the similar absorption edges, the absorptions arising from electronic transitions, can be ignored in the wavelengths range concerned. To distinguish the extinction by scattering effect from that by vibrational absorptions, we analyse the magnitude of different-quanta molecular vibrations from NIR to visible spectral region.

With the wavelength decreasing from 1400 nm to 350 nm (Figure 4), the solvents $H_2O$, $D_2O$, DMSO and DMSO-$d_6$ exhibit that: (1) the extinctions in $H_2O$ and $D_2O$ both drop exponentially till their 5th harmonics, at about 606 and 825 nm, respectively; (2) SO- and CSC-based overtones and combination bands in DMSO are not detectable as these low-lying fundamental vibrations (1041-1052 $cm^{-1}$ for SO, and 667-697 $cm^{-1}$ for CSC) become overcompensated in the wavelengths range concerned; [1] (3) and CH overtones in DMSO decrease rapidly until the 4th harmonic, at around 736 nm. These observations are in good agreement with the general rule of that the intensity of the harmonics decays exponentially as a function of decreasing wavelength, since the vibration intensity is supposed to reduce by about an order of magnitude for every additional quantum number. [1, 33]

Yet, with respect to the extinctions in shorter wavelengths, their intensities commence running counter to the exponentially decreased trend as discussed above: (1) the higher-order vibrations (> the 5th harmonics) in $H_2O$ and $D_2O$ appear overwhelmed and unobservable; (2) the ratio between higher-order harmonics (> the 5th CH overtones) in DMSO no longer holds for being 0.1, e.g., the ratio of its 5th- to 4th-overtones intensity grows to about 0.43; (3) and the ratio of the-same-wavelength extinctions between DMSO and DMSO-$d_6$ shrinks to in between 1.5 and 2.0 from the wavelength of DMSO's 4th CH overtone onwards, standing in contrast to about an order of magnitude difference between their extinctions in the longer wavelengths region. Without the need to consider electronic transitions-induced absorptions, the tendency of the divergence of extinctions from the exponentially decayed harmonics can only be related with the light scattering in solvent, whose intensity, for example, is inversely proportional to the fourth power of the wavelength if that is Rayleigh scattering. With the wavelength decreasing, the scattering strength increases progressively, and its contributions to the extinctions first blend in with that by molecular vibrations and then take the dominance in extinctions. The wavelengths from where the light scattering is considerable or at least comparable with harmonics are nearly at 570, 780, 736 and 1001 nm for solvents $H_2O$, $D_2O$, DMSO, and DMSO-$d_6$, respectively, as shown in Figure 4.

*3.2 Light scattering in solvent: a theoretical and experimental study*

On the basis of the density fluctuation theory, the Einstein-Smoluchowski equation and its two-variant forms are developed and applied to predict the light scattering in solvent. The Einstein-Smoluchowski equation, with the inclusion of the depolarization ratio $\Delta$, is written as: [7, 10]

$$\tilde{b} \ (dB/cm) = \frac{1}{10 Ln 10} \frac{8\pi}{3} R_{tot} \frac{2+\Delta}{1+\Delta} \qquad (7)$$

$$R_{tot} = \frac{\pi^2}{2\lambda_0^4} kT\beta_T \frac{(n^2-1)^2(n^2+2)^2}{9} \frac{6+6\Delta}{6-7\Delta} \qquad (8)$$

where $\tilde{b}$ is the scattering coefficient, $R_{tot}$ the Rayleigh ratio, a measure of the light-scattering power of a medium, $k$ the Boltzmann constant, $T$ the absolute temperature, $\lambda_0$ the wavelength of light in vacuum, $\beta_T$ the isothermal compressibility, $n$ the real part of refractive index at temperature $T$, and $\left.(6+6\Delta)\middle/(6-7\Delta)\right.$ the Cabannes factor. By formulating the density fluctuation at a microscopic level with the pressure derivative of the refractive index at the constant temperature $(\partial n/\partial P)_T$ (so-called isothermal piezo-optic coefficient), Coumou *et al.* and Kratohvil *et al.* modified the equation as follows (first variant): [10, 12, 23]

$$R_{tot} = \frac{2\pi^2}{\lambda_0^4} kTn^2 \frac{1}{\beta_T} \left(\frac{\partial n}{\partial P}\right)_T \frac{6 + 6\Delta}{6 - 7\Delta}. \tag{9}$$

By drawing on the spherical cavity model, Zhang and Hu further amended the Einstein-Smoluchowski equation as (second variant): [11]

$$R_{tot} = \frac{\pi^2}{2\lambda_0^4} kT\beta_T (n^2-1)^2 \left[1 + \frac{2}{3}(n^2+2)\left(\frac{n^2-1}{3n}\right)^2\right]^2 \frac{6 + 6\Delta}{6 - 7\Delta} \tag{10}$$

As manifested in Figure 5a, the Eqs. (8), (9) and (10) have similar capabilities to predict the $H_2O$'s scattering intensity, consistent with our projections based upon the nearly mathematical equality of $2\rho n\left(\frac{dn}{d\rho}\right) = \frac{(n^2-1)(n^2+2)}{3} \approx \frac{2n}{\beta_T}\left(\frac{\partial n}{\partial P}\right)_T \approx (n^2-1)\left[1 + \frac{2}{3}(n^2+2)\left(\frac{n^2-1}{3n}\right)^2\right]$ ($\rho$, the density of solvent). Therefore the choice of the best formula to describe the solvent scattering, at least for $H_2O$, simply matters to the equation's practicality. Though Eq. (9) has been widely used since 1960s [34-37], lack of the data for $\left(\frac{\partial n}{\partial P}\right)_T$ in the most solvents gets its execution bumped. By contrast, the numerical values for terms $n$ and $\beta_T$, both used in Eqs. (8) and (10), are extensively accessible or can be readily measured if they are not published. Hence to theoretically predict the light scattering in solvent, the substitution Eq. (9) with the other two equations is recommended.

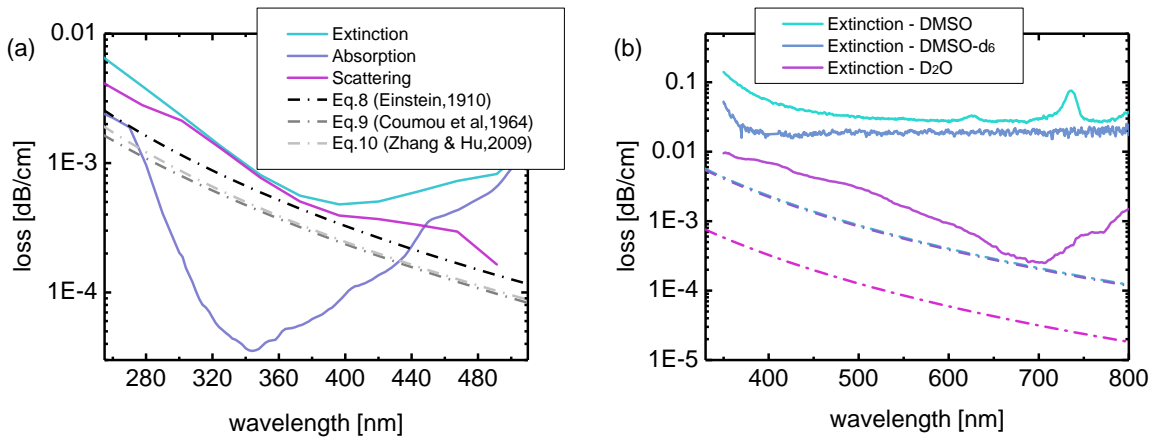

**Figure. 5** (a) The empirical scattering coefficients of $H_2O$ from 255 to 490 nm (pink colour solid line), overlaid with the theoretical results (dash-dot lines). The scattering coefficient is derived by subtracting the absorption coefficient (measured by Mason *et al* [14]) from the extinction coefficient (the interpolated results from Segelstein [28] and Boivin *et al.* [13]). The absorption coefficients of $H_2O$ (light violet solid line) exhibit a dip with minimum at 344 nm, [14] that is a kink point of the respective tails of the solvent's (first) electronic transition and molecular harmonics; and $H_2O$'s scattering intensity is found to account for more than 70% of the solvent extinctions over the wavelengths 275-425 nm. Variable $(\partial n / \partial p)_T$ in Eq. (9) is obtained by linearly extrapolating Kraotohvil *et al.*'s results. [10] (b) The measured extinction coefficients of solvents $D_2O$, DMSO, and DMSO-$d_6$ (solid lines), overlaid with their theoretical scattering coefficients according to the Einstein-Smoluchowski equation (dash-dot lines). Of note, the absorption coefficients for these three solvents are absent in literature, hence their scattering coefficients cannot be extracted as the above did for $H_2O$. The theoretical-curve colours correspond to those for the experimental results. The values of wavelength-insensitive terms $\beta_T$ and $\Delta$ (a weak function of wavelength over the visible range [38]) and wavelength-dependent variable $n$ are exemplified in Table S2.

Figure 5a displays the small disparity existing between $H_2O$'s empirical scattering coefficients and its calculated results (from the Einstein-Smoluchowski equation and its variants), demonstrating their almost identical capacities of the present equations set for predicting the light scattering of $H_2O$. This literally encourages us to examine about how versatile the equations truly are, and for this purpose we look at other common solvents of $D_2O$, DMSO and DMSO-$d_6$, with the aid of the Four-step method. As revealed in Figure 5b, over the wavelengths from 325-800 nm, the theoretical scattering coefficients for these three solvents are lower than those of the measured by factors of about 10 up to 100. As detailed in Table S2, % diff. between the experimental and theoretical results for solvents $D_2O$, DMSO and DMSO-$d_6$ at specified wavelengths show upsurges to 900%, 3553% and 2250%, respectively, against the mild increase of about 30% for $H_2O$. We confirm that these upsurges are irrelative to the act that % diff. are accounted by without subtracting the absorption coefficients from the solvents' extinction coefficients, because even when $H_2O$ absorption portions are included in the analysis, its % diff. just climbs from 30% slightly up to 33%. In addition, we show that such discrepancies persist despite the record-low results for solvents $H_2O$ and $D_2O$, measured by the One-step method, [13] are used to compare with the theoretical predications (Figure S5). The consistent mismatch between the theoretical predications and experimental measures appear to suggest that the present equations are not comprehensive yet for describing the light scattering that could occur to a solvent.

This presumption is validated by the distinct extinction profiles for different solvents: flare groove for DMSO and DMSO-$d_6$, but V-type groove for $H_2O$ and $D_2O$ (Figure 5b). Since the electronic transitions-induced absorptions can be ignored across the visible spectral region for solvents DMSO and DMSO-$d_6$, their extinctions are primarily caused by the intermixing of light scattering and vibrational harmonics. Since the harmonics decay exponentially as the wavelength decreasing (as discussed in Section 3.1), their intensities, in the wavelengths range < 736 nm for DMSO, and < 1001 nm for DMSO-$d_6$, are already orders of magnitude less than those of fundamental resonances, and are unlikely responsible for the prominent extinctions rarely rising or falling over the visible wavelengths in DMSO and DMSO-$d_6$. So this narrows down the solvent scattering responsible for the peculiar flare-groove extinctions, which however immediately conflicts with the predictions by the Einstein-Smoluchowski equation which suggests the intensity of light scattering in solvents varies nonlinearly as a function of the wavelength. As such, we are necessarily lead to point out that the Einstein-Smoluchowski equation is not a complete form to describe the light scattering in solvent, and the full theoretical framework shall at least consider the factors such as molecule size, shape and even trajectory, which then might be able to explain the asymmetric scatterings observed for solvents DMSO/DMSO-$d_6$ and $D_2O$ (Figure 5b).

## 4. Conclusion

In summary, we demonstrate a new spectrophotometry – the Four-step method, achieving sensitivity $10^{-4}$ dB/cm via using 0.5 cm differential path length in standard double-beam spectrophotometre. Our results allow us to position the method as the nearest rival to the record measurement technique, but with reducing the path length by two orders of magnitude, a step-change to conventions from the past 100 years. By comparing the experimental results with the theoretically calculated scattering coefficients, we revitalise the Einstein-Smoluchowski equation for describing the scattering in $H_2O$ due to its high accuracy and simple practicality. With further investigating other solvent objects, we demonstrate that all the present equations have not matured into as a generalised formula to reflect the light scattering taking placing to a solvent, which urges development of an improved theory for better comprehension of the solvent scattering phenomenon.

ASSOCIATED CONTENT

**Supporting Information**

The Supporting Information is available free of charge on the Publications website at DOI: ….

Materials and methods, supporting figures and tables, and additional references (PDF)

AUTHOR INFORMATION

**Notes**

The authors declare no competing financial interests.

ACKNOWLEDGMENT

J.Z. acknowledges the support from the Alexander von Humboldt foundation.

# An improved spectrophotometry tests the Einstein-Smoluchowski equation: a revisit and update

*Jiangbo (Tim) Zhao[†]\*, Cong Qi[†], Guangrui Li[†], and Markus A. Schmidt[†,‡,⊥]*

[†] Leibniz Institute of Photonic Technology, Albert-Einstein-Straße 9, 07745 Jena, Germany

[‡] Abbe Center of Photonic and Faculty of Physics, Friedrich-Schiller-University Jena, Max-Wien-Platz 1, Jena 07743, Germany

[⊥] Otto Schott Institute of Material Research, Fraunhoferstr. 6, 07743 Jena, Germany

E-mail: jiangbo.zhao@leibniz-ipht.de

**Table S1.** List of spectrophotometric measurements of $H_2O$ at room temperature in the near UV to NIR wavelengths range, a complement to the summary over the period 1891-1997 in Ref. [1].

| Year | Method | Instrumental | Differential path length (cm) | Mean % diff. [a] | Refs. |
|------|--------|--------------|------------------------------|------------------|-------|
| 1934 | Single-beam | Customer-designed photographic photometer | 272 | ~ 61 [b] | [2] |
| 1963 | Single-beam | Customer-designed spectrophotometer | 132 | ~ 17 [c] | [3] |
| 1986 | Single-beam | Customer-designed spectrophotometer | 50.3 | ~ -55 [d] | [4] |
| 1999 | Single-beam | Customer-designed spectrophotometer | 150 | ~ 2.1 [e] | [1] |
| 2009 | Double-beam | Commercial spectrophotometer | 0.9 | ~ 9900 | [5] |
| 2012 | Single-beam | Customer-designed spectrophotometer | 100 | ~ 105 [f] | [6] |
| 2015 | Double-beam | NA | 5 | ~ 181 | [7] |
| 2017 | Single-beam | Customer-designed ellipsometer | 20 | ~ 323 | [8] |
| 2019 | Double-beam | Commercial spectrophotometer | 0.5 | ~ -22 | This work |

[a] Mean % diff. = 100% (Exp$_1$-Exp$_2$)/Exp$_2$, where Exp$_{1,2}$ represent the extinction coefficients from the reference as noted, and from the publication by Segelstein [9], respectively. Unless specified, the extinction coefficient refers to the figure at 480 nm.

[b,c] Mean % diff. calculations refer to the extinction coefficients at 400 nm since the figures at 480 nm are absent.

[d] Mean % diff. calculation refers to the extinction coefficient at 436 nm due to the absence of data at 480 nm.

[e] Mean % diff. calculation refers to the extinction coefficient at 400 nm because of negative value at 480 nm.

[f] Mean % diff. calculation refers to the extinction coefficient at 500 nm, the starting wavelength of the measurement.

**Table S2.** Parameters used for calculation of scattering coefficients of solvents $H_2O$, $D_2O$, DMSO, and DMSO-$d_6$. Unless specified, the tabulated data exemplify the figures for $H_2O$ at 344 nm, for $D_2O$ at 650 nm, and for DMSO and DMSO-$d_6$ both at 500 nm. Empirical scattering coefficient $\tilde{b}$ for $H_2O$ and measured extinction coefficients $\tilde{\mu}$ for the other three solvents, are shown in the table, respectively.

| | $\Delta$ | $\beta_T$ ($\times 10^{-10}$) | $\gamma$ ($\times 10^{-10}$) | $n$ [f] | $\tilde{b}\ or\ \tilde{\mu}$ ($\times 10^{-4}$) | $\tilde{b}$ ($\times 10^{-4}$) | % diff. [g] |
|---|---|---|---|---|---|---|---|
| | | m²/N | mN/m | | dB/cm (Exp) | dB/cm (Eq.7&8) | |
| $H_2O$ | 0.108 [a] | 4.58 [10] | 71.98 [11] | 1.35008 | 7.71 | 5.93 | 30 |
| $D_2O$ | 0.111 [b] | 4.74 [12] | 71.87 [13] | 1.33243 | 4.30 | 0.43 | 900 |
| DMSO | 0.436 [c] | 5.32 [10] | 43.54 [14] | 1.48194 | 317.8 | 8.70 | 3553 |
| DMSO-$d_6$ | 0.439 [d] | 5.28 [e] | 43.7 [15] | 1.47801 | 197.6 | 8.41 | 2250 |

[a] The depolarization ratio $\Delta$ for $H_2O$ is the mean of the measured values in the reference. [16]

[b] The net difference of depolarization ratio $\Delta$ between $D_2O$ and $H_2O$ is 0.003, thus yielding 0.111 for $D_2O$. [17]

[c] The depolarization ratio $\Delta$ for DMSO is given by extrapolating the solvent's $\Delta$ in the temperatures range 30-48 °C to the room temperature via a polynomial fit ($R^2$=1). [18] The extrapolated value 0.436 for DMSO is valid, considering 0.453 for the structurally similar solvent Dimethylformamide (DMF). [19]

[d] The depolarization ratio $\Delta$ for DMSO-$d_6$ is about 0.439, obtained by assuming that the net difference in [b] is retained for the pair of solvents DMSO and DMSO-$d_6$.

 The isothermal compressibility $\beta_T$ for DMSO-$d_6$, not found in the literature, is derived on the basis of equation $\beta_T \gamma^{7/4} = $ constant, [20] where $\gamma$ is the surface tension, and "constant" for solvent DMSO is lent to DMSO-$d_6$ because the products $\beta_T \gamma^{7/4}$ between $H_2O$ and $D_2O$ are nearly equal.

f Listing of refractive indices $n$ at 25 °C for the wavelengths as specified in the table caption for the respective solvents. The full data set of refractive indices for different solvents across the wavelengths concerned refer to an index-related Manuscript in preparation by Zhao J, *et al*., for the sake of consistency (factors like material, temperature, etc.) though many other literature have reported the index values.

g % diff. = (Exp-Cal)/Cal, where Exp is the measured coefficient, and Cal represents the calculated results.

## Section S1. Experimental preparations and instrumental evaluation

Accurate measure of extinction coefficients for chemically pure and optically transparent solvent is not as simple and straightforward as it looks. For the sake of reproducibility and rigor in research, we enumerate in detail all of the procedures pertinent to the spectrophotometric measurement.

### S1.1 Solvent

Ultrapure $H_2O$, with resistivity of 18.2 MΩ-cm and total organic carbon (TOC) below 1 ppb, is supplied by a SG Ultra Clear UV plus TM/EI-Ion® system. Deuterated water ($D_2O$, 99.9 atom % D), dimethyl sulfoxide (DMSO for spectroscopy, purity ≥ 99.8 %) and deuterated DMSO (DMSO-$d_6$, 99.9 atom % D) are purchased from Sigma-Aldrich and used without further purification. The solvents are only dispensed before the measurement conduct.

In general, a high-purity solvent, particularly ultrapure $H_2O$, is the corrosive substance, which leaches away present impurities from the parts that the solvent comes into contact with. To minimise this contaminating possibility, we thoroughly cleaned containers such as Pyrex glass vial, PTFE/silicone septum and cuvette, before their use.

**S1.2 Cuvette**

The use of quality and flawless cuvette is entailed, because in addition to being as a sample container, cuvette is automatically a part of the optical system once placed in the compartment. In this work we purchase brand-new Hellma 110-QS Quartz Glass cuvettes (high performance, 200-2500 nm), with 0.5 and 1.0 cm path lengths (tolerance accuracy $\pm$ 0.001 cm). We only qualify the cuvette for the experiment if the one in empty can maintain transmissions above 80 % from 200-2500 nm throughout a complete pre-check (that is the repeated spectrophotometric measurements of the cuvette of interest with its frequent in-/out-placement in the compartment). To avoid the post-incurred defects during the experiment, we handle the cuvettes with great care, including transfer, placement, cleaning, etc.

**S1.3 Cleaning**

The cleaning procedure is carried out as follows: immersing the parts, such as cuvettes, glass vials and PTFE lids, in ~ 2% (v/v) Hellmanex III cleaner solution for about 30 min, followed by ultrasonication for less than 5 min at moderate temperature (to facilitate the cleaning), before rinse in sequence by flowing ethanol, distilled $H_2O$, and ultrapure $H_2O$, and then purging $N_2$ for drying. Note that, to effectively leach away residual impurities, we apply ultrapure $H_2O$ for rinsing lastly; and to minimise the physical adsorption of airborne contaminants and particulates, we either clean the parts right before the use or store them in desiccator immediately after cleaning.

**S1.4 Spectrophotometre**

For the best measurement performance across the wavelengths 350-1400 nm, we use double-beam Spectrophotometers JASCO V-660 (working wavelength 187-900 nm) and V-670 (working wavelength 190-2700 nm) for different spectral regions (see detailed discussions below).

To ensure the thermal equilibrium, we warm up the spectrophotometres for more than 60 mins. To acquire stable baseline characteristic, good signal-to-noise ratio and spectra shape, we set the parameters as: spectra resolution 0.5 nm, scanning speed 100 nm/min, and medium response, with the bandwidths, 2 nm for the wavelengths range < 900 nm and 8 nm for that > 900 nm. We chose the change wavelength 900 nm for the grating and detector in V-670.

**S1.5 Instrumental capacity – baseline flatness and baseline drift**

Strictly speaking, the beams' photon flux densities in the reference and sample compartments in double-beam spectrophotometre are disparate, corresponding to unequal photocurrents upon the detector. To eliminate this disparity, it is usual practice to conduct the baseline correction scan, through which the null current between the beams is accounted, and in that time and in that environment, the "zero absorbance" for the instrument is recognised. Ideally with the baseline correction scan, the subsequent spectrophotometric scans should allow to directly and accurately measure the absorbance of the sample. But the instrument imperfection renders the baseline never unchanged over the time and the wavelengths. In other words, the baseline characteristics, such as baseline flatness (also called instrumental noise) and baseline drift (also known as absorbance stability), would influence the spectrophotometre's capacities, like the accuracy, precision and detection limit of the instrument.

*Baseline flatness (instrumental noise)*

The baseline flatness levels of the two spectrophotometres are specified by the manufacturer, both ± 0.0005 Abs over the wavelengths 200-800 nm. Given that switching of light sources, gratings and detectors could take places over an extended wavelengths range from 350-1400 nm, we reckon that the instrumental noise across the wavelengths of interest likely differ from what the manufacturer has claimed. We thus undertake re-evaluation of the baseline flatness for both spectrophotometres, by means of conduct of six individual "zero absorbance" measurements. Note that, in an attempt to remove interference by baseline drift (as discussed in the subsequent subsection), we perform the baseline correction scan before conducing every "zero absorbance" measurement, i.e., six baseline correction scans in total.

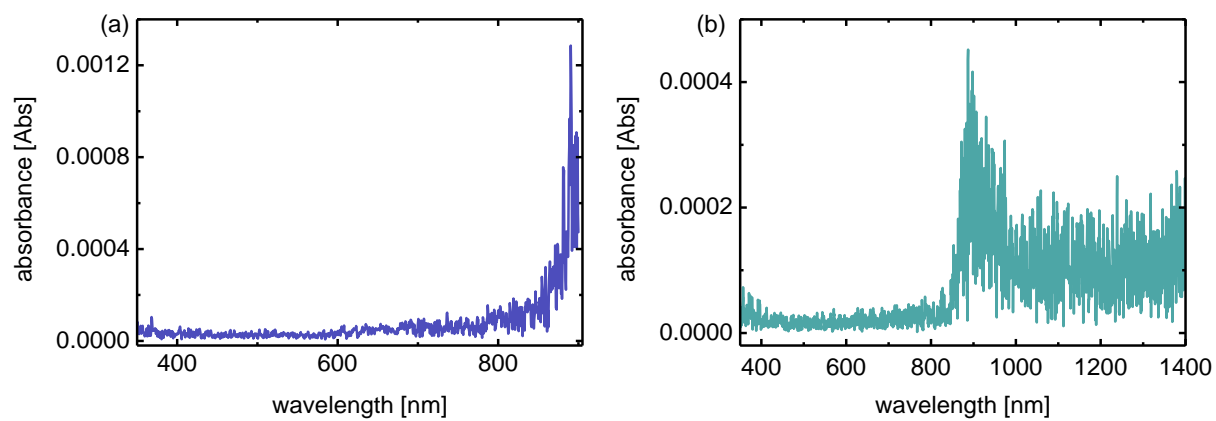

**Figure. S1** The respective instrumental noises of Spectrophotometres V-660 (a) and V-670 (b), obtained by applying the standard deviation (SD) equation to the six "zero absorbance" scans (for a reduced sampling bias).

As shown in Figure S1a, the instrumental noise of V-660 is within ± 0.00003 Abs from 350-650 nm, multiplies in between ± 0.00005 Abs for the range 650-800 nm, and mounts up to ± 0.001 Abs for 800-900 nm, reflecting the suffering sensitivity of photomultiplier tube (PMT) as the wavelength increases. In Figure S1b, it shows the instrumental noise of V-670 stabilises within ± 0.00003 Abs from 350-850 nm, ascends to ± 0.0004 Abs from 850-950 nm (due of the switching of detectors, e.g., from PMT to PbS photoconductive cell, and associated filters), and then falls to ± 0.0002 Abs in the wavelengths range 950-1400 nm. These results indicate that, with regard to the baseline flatness, spectrophotometres V-660 and V-670 are interchangeable.

*Baseline drift (absorbance stability)*

Along with filter switching, light source exchanging, etc., each additional scan would drift the baseline. After multiple scans, the drift cumulates in a way that the instrument's on-going baseline could much differ from the initially recorded baseline. Paradoxically, the resulting absorbance is accounted for by the initial baseline record instead of the real-time baseline. This suggests that the baseline drift should incur the deviation of measured values from actual absorbance, and the drift' degree determines the accuracy level of the measurement. For measurement involving multiple scans over a period of time, evaluation of baseline drift is essential and necessary.

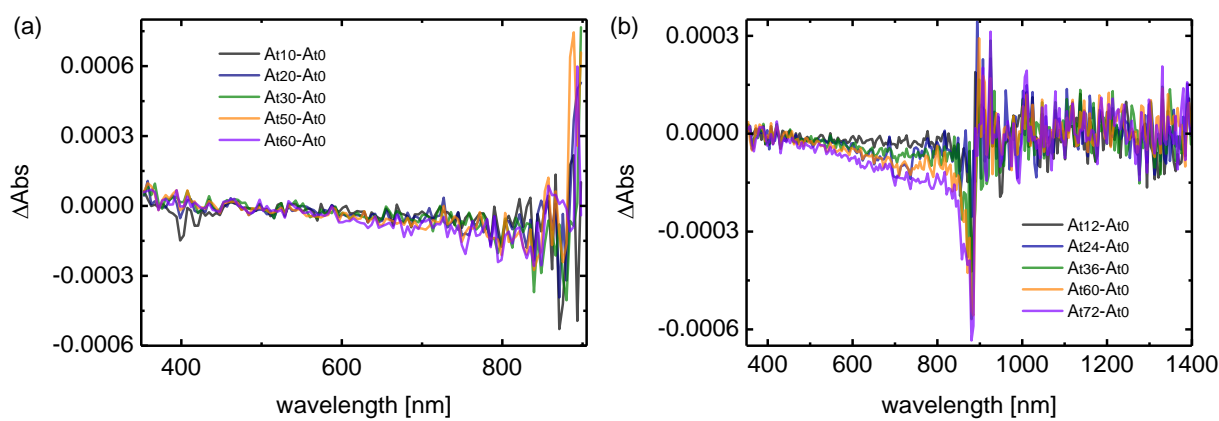

**Figure. S2** Characterisations of baseline drift for Spectrophotometres V-660 (a) and V-670 (b). $|A_{t1} - A_{t2}| = \Delta Abs$, where $A_{t1/2}$ represent the absorbance measured at time $t_1$ and $t_2$ relative to the baseline scan t=0, and $\Delta Abs$ is the net difference in the time intervals as defined. Different from the baseline flatness evaluation where baseline correction scan is needed to be undertaken for every "zero absorbance" measurement, the characterisation of baseline drift only performs baseline correction scan once.

As shown in Figure S2a, the baseline drift for spectrophotometre V-660 is inconsiderable over the wavelengths 350-900 nm in 60 minutes, i.e., $|A_{t0} - A_{t10}| = |A_{t0} - A_{t60}|$. For spectrophotometre V-670, its baseline drift is negligible from 850-1400 nm, while raising progressively in the wavelengths range 350-850 nm, e.g., from less than 0.00001 Abs in 12 min to 0.00005~0.0004

Abs in 72 min (Figure S2b). To access the best accuracy that we could, we decide to use spectrophotometres V-660 and V-670, for 350-850 nm and 850-1400 nm, respectively.

With scanning speed at 100 nm/min, the time of every scan for V-660 (from 350-850 nm) and V-670 (from 850-1400 nm) are about 5 and 5.5 min, respectively. This means that, for the Four-step method measurement, each spectrophotometre spends about 22 min in measuring (for four scans), and as suggested in Figure S2, the baseline drifts are both negligible over this duration.

Defining signal-to-noise ratio (SNR) 2:1 as threshold (2σ), and taking into account baseline flatness and baseline drift, we confirm that the combined use of spectrophotometres offers the detection limit at 0.00006 Abs from 350-850 nm, around 0.0008 Abs from 850-950 nm, and less than 0.0004 Abs from 950-1400 nm.

## Section S2. Spectrophotometric measurement

### S2.1 Beer-Lambert law

According to the Beer-Lambert law, the absorbance ($A$, namely extinction) and transmittance ($T_t$) for a medium are depicted as:

$$A = -Log_{10}\left(\frac{I_t}{I_0}\right) = -Log_{10}T_t = \mu L \qquad (S1)$$

where $I_0$ and $I_t$ are the radiant fluxes of the incident and transmitted light, $\mu$ is the decadic extinction coefficient of the medium (unit: cm$^{-1}$), and $L$ is the path length. The decadic extinction coefficient $\mu$ in unit of cm$^{-1}$ can be converted to the extinction coefficient $\tilde{\mu}$ in unit of dB/cm, via the relation $\tilde{\mu} = 10\mu$. If needed, the napierian extinction coefficient in the literature is converted into its decadic form by multiplying a factor of $1/Ln10$. In this work, we use $\tilde{\mu}$ (or $\mu$) as a measure for the solvent's extinction per unit length.

The extinction coefficient ($\mu$ or $\tilde{\mu}$) is the sum of the absorption coefficient ($a$ or $\tilde{a}$) and scattering coefficient ($b$ or $\tilde{b}$), which is described as:

$$\mu = a + b \text{ or } \tilde{\mu} = \tilde{a} + \tilde{b}. \qquad (S2)$$

*S2.2 Optical paths in cuvettes*

The way a ray of incident light traverses the empty and liquid-filled cuvettes is distinct. Appreciation of their differences in optical paths between the cuvettes is a prerequisite to advance the spectrophotometry.

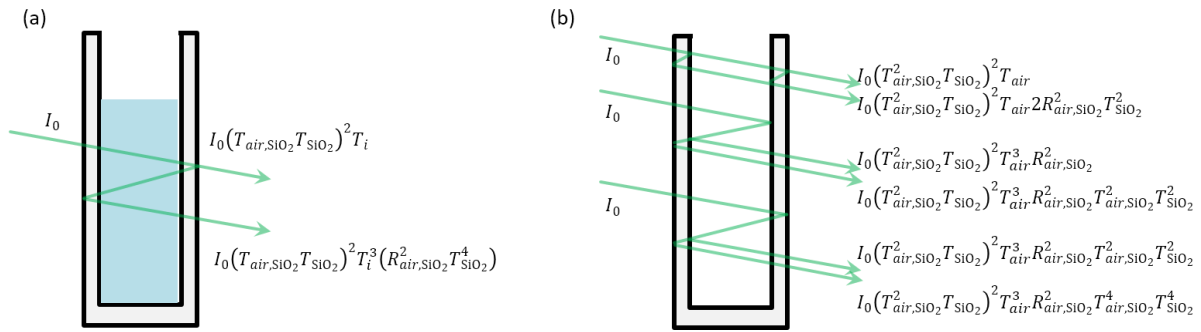

**Figure. S3** Schematic diagrams of the optical paths in the liquid-filled (a) and empty (b) cuvettes. The schematics illustrate the total transmitted light containing the fluxes that have undergone multiple reflections, refractions, and transmissions at the interfaces and over the mediums. For visualization, the normal incidence is sketched intentionally with angle. Given that the thickness of glass wall and the path length of cuvette involved in this work are larger than 0.1 cm (and the spectral bandwidth > 2 nm), the possible light interference effect is neglected. [5]

*Liquid-filled cuvette*

As exemplified in Figure S3a, the total transmitted light $I_{t,f}$ passing through the liquid-filled cuvette is constituted as:

$$I_{t,f} = I_0 \left( T_{air,\text{SiO}_2} T_{\text{SiO}_2} \right)^2 T_i \left( \underbrace{1}_{once\ medium\ trasverse} \right.$$

$$+ \underbrace{R_{air,\text{SiO}_2}^2 T_{\text{SiO}_2}^4 T_i^2}_{thrice\ medium\ trasverses} \qquad \qquad (\text{S3})$$

$$\left. + \underbrace{R_{air,\text{SiO}_2}^4 T_{\text{SiO}_2}^8 T_i^4}_{five\ times\ medium\ trasverses} + \cdots \right)$$

where $R_{i,\text{SiO}_2}$ and $T_{i,\text{SiO}_2}$ are the reflection and transmission for an interface between the medium (either the air or a specific solvent) and cuvette, and $T_{\text{SiO}_2}$ and $T_i$ are the transmissions for the cuvette glass wall and for the medium in cuvette, respectively. The reflections and transmissions, for an interface, are described by Fresnel equations $R_{i,\text{SiO}_2} = \left( \frac{n_{\text{SiO}_2} - n_i}{n_{\text{SiO}_2} + n_i} \right)^2$ and $T_{i,\text{SiO}_2} = 1 - R_{i,\text{SiO}_2}$, where $n_{\text{SiO}_2}$ and $n_i$ are the refractive indices of the cuvette and medium, respectively. The transmissions of the quartz glass and medium are described as $T_{\text{SiO}_2} = 10^{-\mu_{\text{SiO}_2} L_{\text{SiO}_2}}$ and $T_i = 10^{-\mu_i L_i}$, where $\mu_{\text{SiO}_2}$ (and $\mu_i$) and $L_{\text{SiO}_2}$ (and $L_i$) are the extinction coefficients and path lengths.

In the wavelengths range concerned, $R_{air,\text{SiO}_2}$ is less than 0.037, $T_{air}$ for air and $T_{\text{SiO}_2}$ for glass walls both approximately equal to 1, and $T_i$ always less than 1, hence the sum of the geometric series in Eq. (S3) becomes:

$$I_{t,f} = I_0 \left( T_{air,\text{SiO}_2} T_{\text{SiO}_2} \right)^2 T_i (1 + \delta_f) \qquad \qquad (\text{S4})$$

where $\delta_f$, the ratio of a total amount of the higher-order transmitted light to the first-order counterpart, is in the approximation to $0.001 T_i^2$. This is the Eq. (1) in the main text.

*Empty cuvette*

As shown in Figure S3b, the higher-order transmitted light for an empty cuvette consists of the light that passes through the medium once, thrice, and so on, and air-glass interface for six, eight times, etc. Including the first-order transmitted light, the total transmitted light is summed as:

$$I_{t,e} = I_0\left(T_{air,\mathrm{SiO_2}}^2 T_{\mathrm{SiO_2}}\right)^2 T_{air}\big(\underbrace{1 + 2R_{air,\mathrm{SiO_2}}^2 T_{\mathrm{SiO_2}}^2}_{once\ medium\ trasverse} +$$

$$\underbrace{R_{air,\mathrm{SiO_2}}^2 T_{air}^2 + 2R_{air,\mathrm{SiO_2}}^2 T_{air,\mathrm{SiO_2}}^2 T_{\mathrm{SiO_2}}^2 T_{air}^2 + R_{air,\mathrm{SiO_2}}^2 T_{air,\mathrm{SiO_2}}^4 T_{\mathrm{SiO_2}}^4 T_{air}^2}_{thrice\ medium\ trasverses} + \cdots\big). \tag{S5}$$

The summed amount of this infinite series is further simplified to:

$$I_{t,e} = I_0\left(T_{air,\mathrm{SiO_2}}^2 T_{\mathrm{SiO_2}}\right)^2 T_{air}(1 + \delta_e), \tag{S6}$$

where $\delta_e$, the ratio of a total amount of the higher-order transmitted light to the first-order counterpart, lies within 0.0063~0.0078. This is the Eq. (2) in the main text.

Albeit higher-order transmissions, such as $I_0\left(T_{air,\mathrm{SiO_2}} T_{\mathrm{SiO_2}}\right)^2 T_i \delta_f$ and $I_0\left(T_{air,\mathrm{SiO_2}}^2 T_{\mathrm{SiO_2}}\right)^2 T_{air} \delta_e$, in minor quantity, the direct omissions of these terms from Eqs. (S4) and (S6) are not recommended, as their presences are helpful to stress the criterion that a pair of cuvettes used during the measurement must be as much clean, quality, and identical as possible.

## Section S3. Four-step method

### S3.1 Experimental operations

As named, the Four-step method includes four spectrophotometric scans for two-different-length cuvettes in the sample compartment. The operation is specified as follows in order: making the baseline correction scan → performing the spectrophotometric scan for short-length cuvette in empty (Step 1) → filling the liquid of interest in the cuvette followed by running another spectrophotometric scan (Step 2) → replacing with a long-length empty cuvette and then conducting the scan (Step 3) → undertaking the last scan for the liquid-filled long-length cuvette (Step 4). Importantly, to minimise the experimental uncertainties, any inserted cuvettes are kept immobile without touching until the cuvette replacement.

*S3.2 Theoretical derivation*

The incident light $I_0$, dispensed from a stable lamp source under a DC power, is supposed to be invariant. The selected cuvettes after pre-check are viewed as nearly equal, i.e., $n_{\mathrm{SiO_2}}$ and $T_{\mathrm{SiO_2}}$ are almost the same between the cuvettes. As long as these conditions are simultaneously met (so that unknown terms linked to the cuvettes are eliminable), the substitution of the incident light $I_0$ and the transmitted light $I_{L1,e,s}$, $I_{L1,f,s}$, $I_{L2,e,s}$, and $I_{L2,f,s}$ in Eqs. (S4) and (S6) into Eq. (S1) gives:

$$
\begin{aligned}
\Delta A_{L1,s} &= A_{L1,f,s} - A_{L1,e,s} \\
&= -Log_{10}(\frac{T_{L1,f,s}}{T_{air,\mathrm{SiO_2}}^2 T_{air}}\frac{1 + \delta_{L1,f,s}}{1 + \delta_{L1,e,s}})
\end{aligned}
\tag{S7}
$$

and

$$
\begin{aligned}
\Delta A_{L2,s} &= A_{L2,f,s} - A_{L2,e,s} \\
&= -Log_{10}(\frac{T_{L2,f,s}}{T_{air,\mathrm{SiO_2}}^2 T_{air}}\frac{1 + \delta_{L2,f,s}}{1 + \delta_{L2,e,s}}).
\end{aligned}
\tag{S8}
$$

Subtraction between Eqs. (S8) and (S7) derives $\mu$, as:

$$\mu = \frac{\Delta A_{L2,s} - \Delta A_{L1,s}}{L_2 - L_1} + \frac{Y}{L_2 - L_1} \tag{S9}$$

and $Y$ is described as:

$$Y = Log_{10}\left(\frac{1 + \delta_{L2,f,s}}{1 + \delta_{L1,f,s}} \frac{1 + \delta_{L1,e,s}}{1 + \delta_{L2,e,s}}\right) = Log_{10}\left(\frac{1 + \delta_{L2,f,s}}{1 + \delta_{L1,f,s}}\right). \tag{S10}$$

Eqs. (S9) and (S10), i.e., Eqs. (3) and (4) in the main text, are the mathematical expression of the Four-step method. The term $Y$ describes the contribution from the higher-order transmitted light, where, for different-length cuvettes (assumed identical), the glass walls hold the equality between $\delta_{L1,e,s}$ and $\delta_{L2,e,s}$, but the different lengths lead to the disparity between $\delta_{L2,f,s}$ and $\delta_{L1,f,s}$, as indicated in Eqs. (S3) and (S5). Given the solvent's extinction coefficients from 0.00001 to 0.1 cm$^{-1}$ ($L_1$=0.5 cm and $L_2$=1.0 cm), the term $Y/(L_2 - L_1)$ varies between -2.0·10$^{-8}$ and -1.4·10$^{-4}$ cm$^{-1}$, corresponding to errors from -0.20 % to -0.14 %. This $Y$ is inconsiderable and neglected during the execution of Eq. (S9), but its retention in the formula is used to emphasize the criterion that the cuvettes must be as identical as possible.

*S3.3 Theoretical derivation of the Two-step method*

Similar to the derivation process for the Four-step method, formulation of the "Two-step method" for solvent extinct coefficient is written as:

$$\mu = \frac{A_f - A_e}{L_{2,s} - L_{1,r}} + 2Log_{10}\left(\frac{T_{air,SiO_2,L2,s}}{T_{air,SiO_2,L1,r}}\right) + \frac{Y}{L_{2,s} - L_{1,r}}, \tag{S11}$$

where $A_f = Log_{10}\frac{I_{t,f,L1,r}}{I_{t,f,L2,s}}$, $A_e = Log_{10}\frac{I_{t,e,L1,r}}{I_{t,e,L2,s}}$, and $Y$ is described as:

$$Y = Log_{10}\left(\frac{1 + \delta_{L1,f,r}}{1 + \delta_{L2,f,s}} \frac{1 + \delta_{L2,e,s}}{1 + \delta_{L1,e,r}}\right). \tag{S12}$$

The beams between the sample ($s$) and reference ($r$) channels are different, hence the second and third items in Eq. (S11) cannot be eliminated. Any cuvettes are imperfect, such that the differences between the two items are further amplified. This accounts for the large errors encountered when using the Two-step method, because omission of the last two items in Eq. (S11), as a majority of measurements exercised to date, makes the extinction coefficient measured largely deviate from the true value.

## Section S4. Compilation of measurement uncertainties

### S4.1 Theoretical compilation of measurement uncertainties

For the Four-step method, extraction of solvent extinction coefficient relies on Eq. (S9). A propagation of error of the variables in Eq. (S9) yields the extinction coefficient uncertainty, expressed as:

$$S_\mu^2 = \left(\frac{1}{L2-L1}\right)^2 S_{(\Delta A_{L2,s}-\Delta A_{L1,s})}^2$$
$$+ \frac{(\Delta A_{L2,s}-\Delta A_{L1,s})^2}{(L2-L1)^4} S_{(L2-L1)}^2 \qquad (S13)$$
$$- 2\frac{\Delta A_{L2,s}-\Delta A_{L1,s}}{(L2-L1)^3} S_{AL}^2,$$

where

$$S_{(\Delta A_{L2,s}-\Delta A_{L1,s})}^2 = S_{A_{L2,f,s}}^2 + S_{A_{L2,e,s}}^2 + S_{A_{L1,f,s}}^2 + S_{A_{L1,e,s}}^2 \qquad (S14)$$

$$S_{(L2-L1)}^2 = S_{L2}^2 + S_{L1}^2. \qquad (S15)$$

The cross term $S_{AL}$ is not given since its sum approaches zero.

The standard deviation of the absorbance, $S_A$, is described as:

$$S_A^2 = \left(\frac{\partial A}{\partial L}\right)^2 S_L^2 + \left(\frac{\partial A}{\partial \theta}\right)^2 S_\theta^2 + \left(\frac{\partial A}{\partial n_{SiO_2}}\right)^2 S_{n_{SiO_2}}^2, \quad\quad (S16)$$

where $S_L$, $S_\theta$ and $S_{n_{SiO_2}}$ are the standard deviations for the corresponding variables, and possible covariances between $\theta$, $L$ and $n$ are negligible, and

$$\frac{\partial A}{\partial L} = \mu \quad\quad (S17)$$

$$\frac{\partial A}{\partial n_{SiO_2}} = \left(\frac{1}{n_{SiO_2}} - \frac{1}{n_{SiO_2}+1} - \frac{1}{n_{SiO_2}+n_i}\right)\frac{4}{\ln 10}. \quad\quad (S18)$$

The full expression $\frac{\partial A}{\partial \theta}$ is a function of $sin\,\theta$, thus $\lim_{\theta \to 0}\frac{\partial A}{\partial \theta} = 0$, suggesting the error from the small uncertainty of $\theta$ is not consequential. Substitution of Eqs. (S14-S18) into Eq. (S13) yields the SD for extinction coefficient as:

$$S_\mu = \frac{1}{L2-L1}\sqrt{6\mu^2 S_L^2 + \frac{64}{(\ln 10)^2}\left(\frac{1}{n_{SiO_2}} - \frac{1}{n_{SiO_2}+1} - \frac{1}{n_{SiO_2}+n_i}\right)^2 S_n^2}. \quad\quad (S19)$$

The variables concerned here are not intended to be exhaustive. For example the insignificant temperature fluctuation is not considered. As suggested by Højerslev and Trabjerg [21] and Röttgers et al., [22] the temperature-dependent offset is at the level of $10^{-7}$ cm$^{-1}$ °C$^{-1}$, which is ignorable against the aforesaid variables' uncertainties.

*S4.2 Experimental validation of the measurement limit*

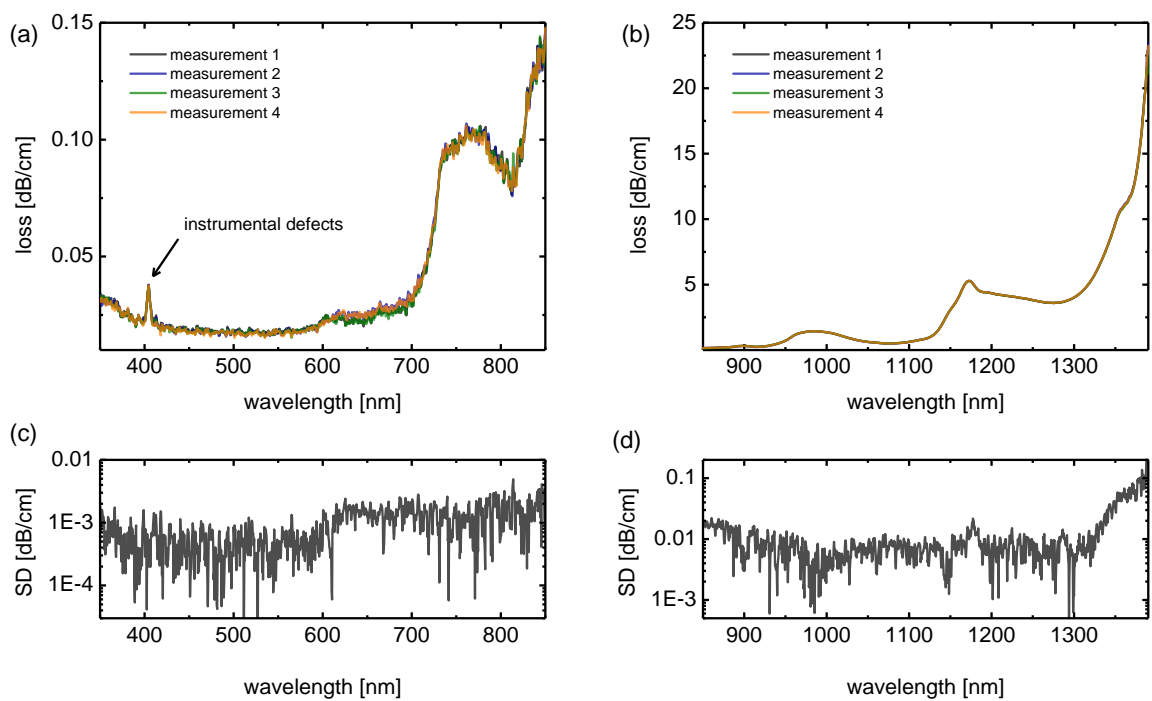

**Figure. S4** The Four-step method-measured extinctions in $H_2O$-DMSO mixture (with mass ratio 70:30), over (a) 350-850 nm for V-660, (b) and 850-1400 nm for V-670 (×4 measurements, equivalent to a total of ×16 scans for each spectrophotometre). (c, d) The SD of extinction coefficients in the mean of ×16 scans for the corresponding spectrophotometres, in good agreement with the detection limit discussed in Section S1.5.

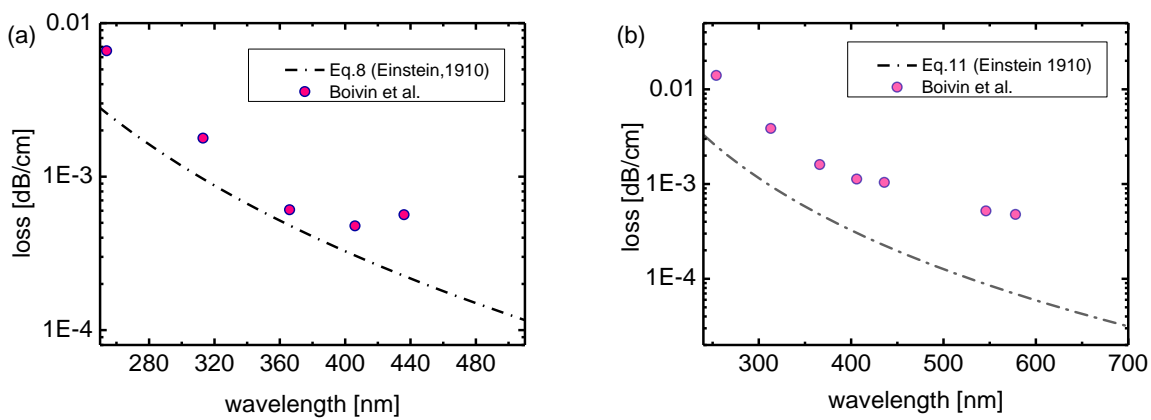

**Figure. S5** The extinction coefficients for solvents (a) $H_2O$ and (b) $D_2O$, where the record-low extinctions (filled circles) measured by the One-step method,[4] are overlaid with the theoretical predications according to the Einstein-Smoluchowski equation (dash-dot lines).

## Section S5. Summary of fundamental vibration frequencies in solvents

Solvent $H_2O$ exhibits three fundamental vibrations associated with two OH bonds, including symmetric ($\upsilon_1$) and asymmetric ($\upsilon_3$) stretch, and bending ($\upsilon_2$), with resonances maxima of approximately 3277 cm$^{-1}$ (3.05 μm), 3490 cm$^{-1}$ (2.87 μm) and 1645 cm$^{-1}$ (6.08 μm). Solvent $H_2O$'s absorptions around 606, 739 and 970 nm are assigned to the respective $n$th harmonics of stretching, and 1200 nm resonance peak indexed to the combination band of 2nd overtone of stretching and fundamental bending. [23-26]

For DMSO, in addition to the weak rocking vibrations of $CH_3$ at 1019 and 1031 cm$^{-1}$, the solvent gives rise to multiple prominent broad resonances, such as the fundamental stretching vibrations from $CH_3$, $CH_2$, & CH in range of 2850-3000 cm$^{-1}$ (3.33-3.50 μm), the fundamental bending resonances from $CH_2$&$CH_3$-deformations between 1350-1470 cm$^{-1}$ (6.8-7.4 μm), the SO fundamental stretching modes from 1041-1052 cm$^{-1}$ (9.5-9.6 μm), and the CSC fundamental stretching vibrations from 667-697 cm$^{-1}$ (14.3-15.0 μm). DMSO's characteristic bands with maxima at 627, 736, 901 and 1180 nm and a span of 680-740, 756-822, 850-925, 971-1057, 1133-1233 and 1360-1400 nm arise from the overtones of $CH_{1,2,3}$-stretching vibrations (2-9th overtones) and $CH_2$&$CH_3$-deformations (4-9th overtones), respectively. [27-31]